\documentclass[preprint,onecolumn]{revtex4}
\usepackage[pdftex]{graphicx}

\newcommand\ee{\end{equation}}
\newcommand\be{\begin{equation}}

\usepackage{physics}
\makeatother

\begin{document}
\begin{flushright}
MI-TH-1724
\end{flushright}
\title{Non-standard interactions of solar neutrinos in dark matter experiments}

\author{Bhaskar Dutta}
\author{Shu Liao}
\author{Louis E. Strigari}
\affiliation{Mitchell Institute for Fundamental Physics and Astronomy,  Department of Physics and Astronomy, Texas A\&M University, College Station, TX 77845, USA}
\author{Joel W. Walker}
\affiliation{Department of Physics, Sam Houston State University, Huntsville, TX 77341, USA}

\begin{abstract}
Non-standard neutrino interactions (NSI) affect both their propagation through matter and their detection, with bounds on NSI parameters coming from various astrophysical and terrestrial neutrino experiments. In this paper, we show that NSI can be probed in future direct dark matter detection experiments through both elastic neutrino-electron scattering and coherent neutrino-nucleus scattering, and that these channels provide complementary probes of NSI. We show NSI can increase the event rate due to solar neutrinos, with a sharp increase for lower nuclear recoil energy thresholds that are within reach for upcoming detectors. We also identify an interference range of NSI parameters for which the rate is reduced by approximately 40\%. Finally, we show that the ``dark side" solution for the solar neutrino mixing angle may be discovered at forthcoming direct detection experiments.
\end{abstract}

\maketitle

\section{Introduction}
\par Direct searches for dark matter now have strong limits on the spin-independent WIMP-nucleus scattering cross section~\cite{Aprile:2012nq,Akerib:2015rjg,Agnese:2015nto,Angloher:2015ewa,Armengaud:2016cvl}. Future searches that aim to improve upon these limits, or even detect dark matter, will be challenged by neutrino backgrounds from the sun, supernovae, and atmosphere~\cite{Monroe:2007xp,Strigari:2009bq,Billard:2013qya,Ruppin:2014bra}. Thinking about the astrophysical neutrinos as a signal rather than background, future experiments may be able to probe exotic properties of neutrinos or astrophysical properties of the sources~\cite{Pospelov:2011ha,Harnik:2012ni,Billard:2014yka}, and are complementary to terrestrial searches for exotic neutrino interactions~\cite{Cerdeno:2016sfi,Dent:2016wcr,Kosmas:2017zbh,Lindner:2016wff,Shoemaker:2017lzs}.

\par Non-standard neutrino interactions (NSI) describe beyond the Standard Model (BSM) couplings of neutrinos~\cite{Ohlsson:2012kf,Miranda:2015dra}. NSI affect neutrino production, propagation, and detection, and have been searched for through each of these channels. For propagation, the presence of NSI modifies the matter potential through both the diagonal and off-diagonal elements in the effective Hamiltonian. For detection, NSI affects the interactions with leptons or quarks, either enhancing or decreasing the cross section relative to the SM value. Non-standard neutrino interactions have been invoked to explain recent discrepancies in measurements of neutrino masses and mixing angles~\cite{Maltoni:2015kca}, and may be probed by future reactors and accelerators~\cite{Fukasawa:2016nwn,deGouvea:2015ndi}, and through astrophysics~\cite{Friedland:2004ah,Bolanos:2008km,Agarwalla:2012wf,Stapleford:2016jgz,Gonzalez-Garcia:2016gpq}. Though there is not yet evidence for NSI, for some combinations of couplings between fermions and neutrino flavors the bounds are weak, still allowing for couplings of order unity.

\par In this paper, we study NSI of solar neutrinos in dark matter detectors. We focus on solar neutrinos because their flux has been well studied and their interaction rates can be readily compared to the corresponding rates deduced from previous experiments. We consider interactions between all types of neutrino flavors and fermions, and use a three-flavor formalism accounting for NSI in propagation through the solar interior and in detection on Earth. For detection, we consider both elastic neutrino-electron scattering and coherent neutrino-nucleus scattering, and show that these channels are complementary, probing distinct regimes of the NSI parameter space.

\par We identify a range of NSI parameter space that is not ruled out by neutrino experiments, but is observable in dark matter experiments. For certain parameters, we show that due to NSI the event rate can be either enhanced or decreased relative to the SM value. For rates which are increased, we identify parameter ranges that can be probed by forthcoming ton-scale direct dark matter detection experiments~\cite{Aprile:2015uzo,Mount:2017qzi}. We identify an interference range of NSI parameters for which the rate is reduced by approximately 40\%. We additionally show that the ``dark side" solution for solar neutrino mixing angles can be probed by forthcoming dark matter experiments.

\section{Solar neutrinos and Non-standard interactions}

\par For neutral current NSI, the most general four fermion interaction is
\begin{equation}
\mathcal{L}_{int}=2\sqrt{2}G_{F}\bar{\nu}_{\alpha L}\gamma^{\mu}\nu_{\beta L}\left(\epsilon_{\alpha\beta}^{fL}\bar{f}_{L}\gamma_{\mu}f_{L}+\epsilon_{\alpha\beta}^{fR}\bar{f}_{R}\gamma_{\mu}f_{R}\right),
\end{equation}
where $\alpha$, $\beta = e, \mu, \tau$ indicates the neutrino flavor, and $L$, $R$ denote left and right-handed components. From this, the cross section for the interaction between a neutrino and a fermion, $\nu_{\beta}+f\rightarrow\nu_{\alpha}+f$, as a function of nuclear recoil energy, $E_r$, is
\begin{equation}
\dv{\sigma}{E_{r}}=\frac{2}{\pi}G_{F}^{2}m_{f}
\left[\abs{\epsilon_{\alpha\beta}^{fL}}^{2}+\abs{\epsilon_{\alpha\beta}^{fR}}^{2}\left(1-\frac{E_{r}}{E_{\nu}}\right)^{2}
-\frac{1}{2}\left(\epsilon_{\alpha\beta}^{fL*}\epsilon_{\alpha\beta}^{fR}+\epsilon_{\alpha\beta}^{fL}\epsilon_{\alpha\beta}^{fR*}\right)\frac{m_{f}E_{r}}{E_{\nu}^{2}}\right],
\label{eq:crossSect}
\end{equation}
where $m_f$ is the mass of the electron or nucleus~\cite{Barranco:2005yy}. Note that a change of neutrino flavor may be induced by NSI. The $\epsilon$'s of electron scattering in Equation~\ref{eq:crossSect} can be written as
\begin{align}
\epsilon_{\alpha\alpha}^{eL} & \rightarrow\delta_{\alpha e}+\left(-\frac{1}{2}+\sin^{2}\theta_{w}\right)+\epsilon_{\alpha\alpha}^{eL}\\
\epsilon_{\alpha\alpha}^{eR} & \rightarrow\sin^{2}\theta_{w}+\epsilon_{\alpha\alpha}^{eR},
\end{align}
where the NSI contributions are given by the last term on the right hand side of both of these equations, and the remaining terms are SM contributions.

\par Accounting for the spin-up and spin-down components in a nucleus, it is more convenient to use vector and axial vector parameters
$\epsilon^{V}=\epsilon^{L}+\epsilon^{R}$ and $\epsilon^{V}=\epsilon^{L}-\epsilon^{R}$. Then after a short summation,
\begin{align}
\epsilon_{\alpha\alpha}^{L} & \rightarrow\frac{1}{2}Z\epsilon_{\alpha\alpha}^{pV}+\frac{1}{2}\left(Z_{+}-Z_{-}\right)\epsilon_{\alpha\alpha}^{pA}+\frac{1}{2}N\epsilon_{\alpha\alpha}^{nV}+\frac{1}{2}\left(N_{+}-N_{-}\right)\epsilon_{\alpha\alpha}^{nA}\\
\epsilon_{\alpha\alpha}^{R} & \rightarrow\frac{1}{2}Z\epsilon_{\alpha\alpha}^{pV}-\frac{1}{2}\left(Z_{+}-Z_{-}\right)\epsilon_{\alpha\alpha}^{pA}+\frac{1}{2}N\epsilon_{\alpha\alpha}^{nV}-\frac{1}{2}\left(N_{+}-N_{-}\right)\epsilon_{\alpha\alpha}^{nA}
\end{align}
where $Z_{+}$($N_{+}$) and $Z_{-}$($N_{-}$) are the corresponding numbers of spin-up and spin-down protons (neutrons), and the $\epsilon^{p}$ and $\epsilon^{n}$ are
\begin{align}
\epsilon_{\alpha\alpha}^{pV} & =\frac{1}{2}-2\sin^{2}\theta_{w}+2\epsilon_{\alpha\alpha}^{uV}+\epsilon_{\alpha\alpha}^{dV}\\
\epsilon_{\alpha\alpha}^{pA} & =\frac{1}{2}+2\epsilon_{\alpha\alpha}^{uA}+\epsilon_{\alpha\alpha}^{dA}\\
\epsilon_{\alpha\alpha}^{nV} & =-\frac{1}{2}+\epsilon_{\alpha\alpha}^{uV}+2\epsilon_{\alpha\alpha}^{dV}\\
\epsilon_{\alpha\alpha}^{pA} & =-\frac{1}{2}+\epsilon_{\alpha\alpha}^{uA}+2\epsilon_{\alpha\alpha}^{dA}.
\end{align}

\par The structure of the nucleus is described by the quantity $F^{2}\left(Q^{2}\right)$, which we parameterize as the helm form factor~\cite{Lewin:1995rx}. It is also worth noticing that the axial vector contribution is negligible since it is proportional to the nuclear spin, and since the nucleus is heavy, the entire process is non-relativistic, $E_{r}\ll E_{\nu}$.

\par The propagation of neutrinos is described by
\begin{equation}
i\dv{t}\ket{\nu_{\beta}}=\mathcal{H}_{\beta\alpha}\ket{\nu_{\alpha}}
\end{equation}
where the Hamiltonian includes both vacuum and matter contributions
\begin{equation}
\mathcal{H}_{\beta\alpha}=\left[U\mathrm{diag}\left(0,\frac{\Delta m_{21}^{2}}{2E},\frac{\Delta m_{31}^{2}}{2E}\right)U^{\dagger}\right]_{\beta\alpha}
+\sqrt{2}G_{F}\sum_{f}n_{f}\left(\delta^{ef}\delta_{e\alpha}+\epsilon_{\beta\alpha}^{f}\right),
\label{eq:hamoltonian}
\end{equation}
where $U$ is the neutrino mixing matrix and $\Delta m_{\imath \jmath}^2$ are the squared differences between neutrino mass eigenstates. Including only the effect of forward scattering, in which zero momentum is transferred, the NSI parameter $\epsilon$ in Equation~\ref{eq:hamoltonian} is given by $\epsilon_{\alpha\beta}^{f}=\epsilon_{\alpha\beta}^{fL}+\epsilon_{\alpha\beta}^{fR}$. In addition to $\epsilon_{\alpha\beta}^{fL}+\epsilon_{\alpha\beta}^{fR}$, the quantity $\epsilon_{\alpha\beta}^{fL}-\epsilon_{\alpha\beta}^{fR}$ also plays role in the scattering experiments. For the solar density, $n_f$, we take the model of Ref.~\cite{Bahcall:2004pz}.


\par We determine the neutrino survival probability in a full three-flavor framework. The survival probability is obtained by diagonalizing the Hamiltonian
\begin{equation}
\mathcal{H}\left(t\right)=\tilde{U}\left(t\right)\mathrm{diag}\left(\frac{m_{1}^2\left(t\right)}{2E},\frac{m_{2}^2\left(t\right)}{2E},\frac{m_{3}^2\left(t\right)}{2E}\right)\tilde{U}\left(t\right)^{\dagger}.
\end{equation}
The density in the sun changes smoothly enough that neutrinos propagate adiabatically. This means that by averaging over the distance,  the probability of transition from flavor $\beta$ to flavor $\alpha$ is
\begin{equation}
P_{\beta\rightarrow\alpha}=\abs{\tilde{U}\left(t\right)_{\alpha i}}^{2}\abs{\tilde{U}\left(0\right)_{\beta i}}^{2}.
\label{eq:probability}
\end{equation}
This can be interpreted as the multiplication of the probability of transition to a mass eigenstate at the production region and at the escaping region~\cite{Kuo:1989qe}.


We note that previous authors have computed the solar neutrino survival probability including NSI within a reduced two-flavor framework~\cite{Friedland:2004pp}.
We find that the two-flavor framework is a good approximation to the full three-flavor framework, with the only discrepancy arising at high neutrino energy, where the approximation that $G_{F}\sum_{f}n_{f}\epsilon_{\alpha\beta}^{f}\ll\Delta m_{31}^{2}/E_{\nu}$ breaks down. In the energy regime where most solar neutrinos lie (eg. $E_{\nu}<20\,\mathrm{MeV}$), the three-flavor survival probability in Equation~\ref{eq:probability} will give the same result as the more commonly used two-flavor survival probability. The advantage of adopting the three-flavor oscillation framework is that it enables the examination of entire space of $\epsilon$ using scattering experiments.

\section{Results}
\par We identify regimes of the NSI parameter space that are not ruled out by previous experiments, but are observable in direct detection experiments. We take the range quoted in Refs.~\cite{Gonzalez-Garcia:2013usa,Coloma:2017egw} for the allowed range of $\epsilon$'s. In order to isolate the impact of individual NSI parameters, we allow a given $\epsilon$ to vary one at a time, while keeping all others fixed. For simplicity we just present results for a xenon target, though the salient points of our argument are not affected by this choice. Unless otherwise indicated we take the LMA solution for the neutrino oscillation parameters~\cite{Olive:2016xmw}. For the solar neutrino fluxes, we take the high metallicity standard solar model~\cite{Antonelli:2012qu,Robertson:2012ib}, and we include all the components of the solar spectrum.


\par
\begin{figure*}
\begin{tabular}{cccc}
\includegraphics[height=3.5cm]{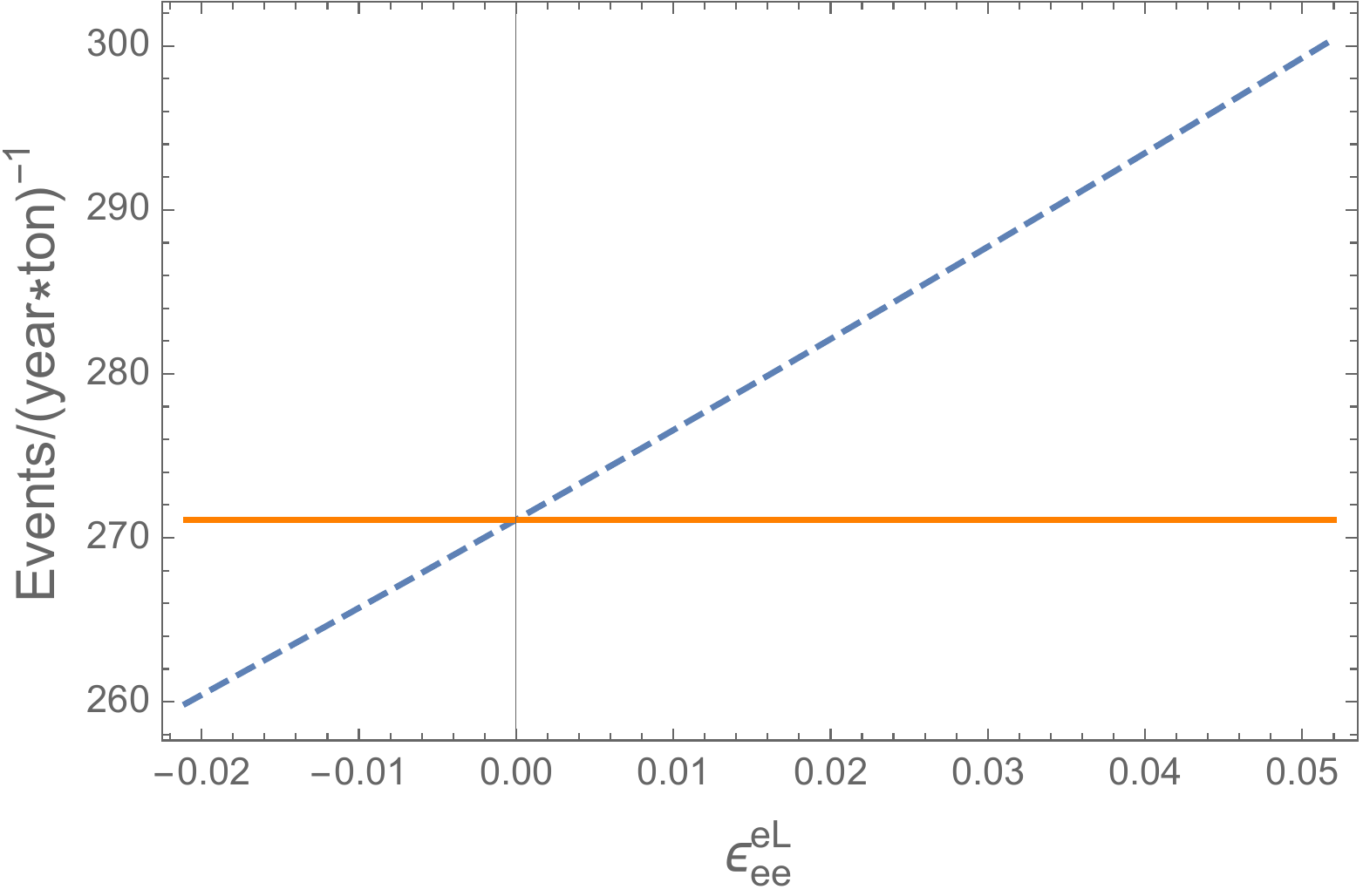} &
\includegraphics[height=3.5cm]{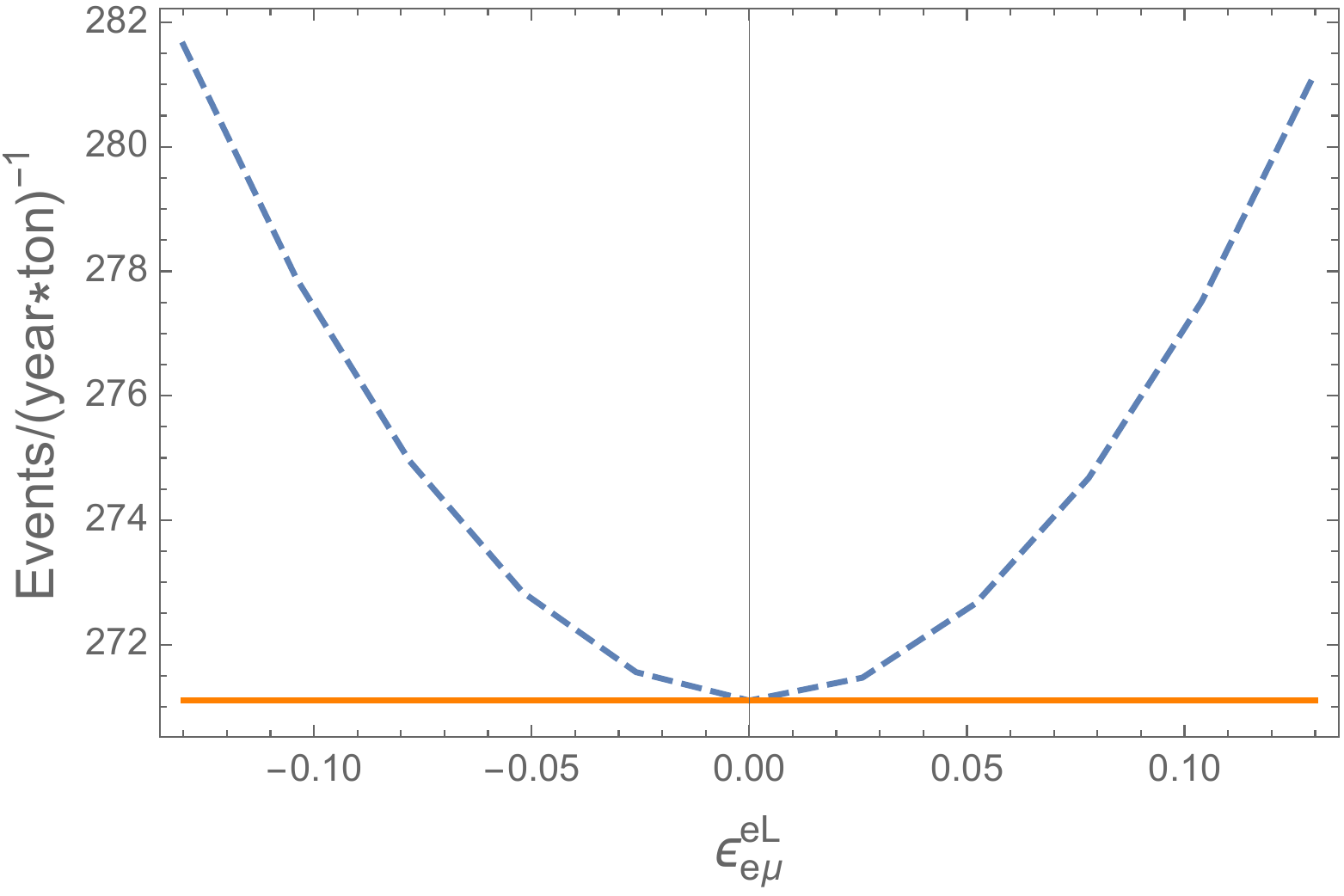} &
\includegraphics[height=3.5cm]{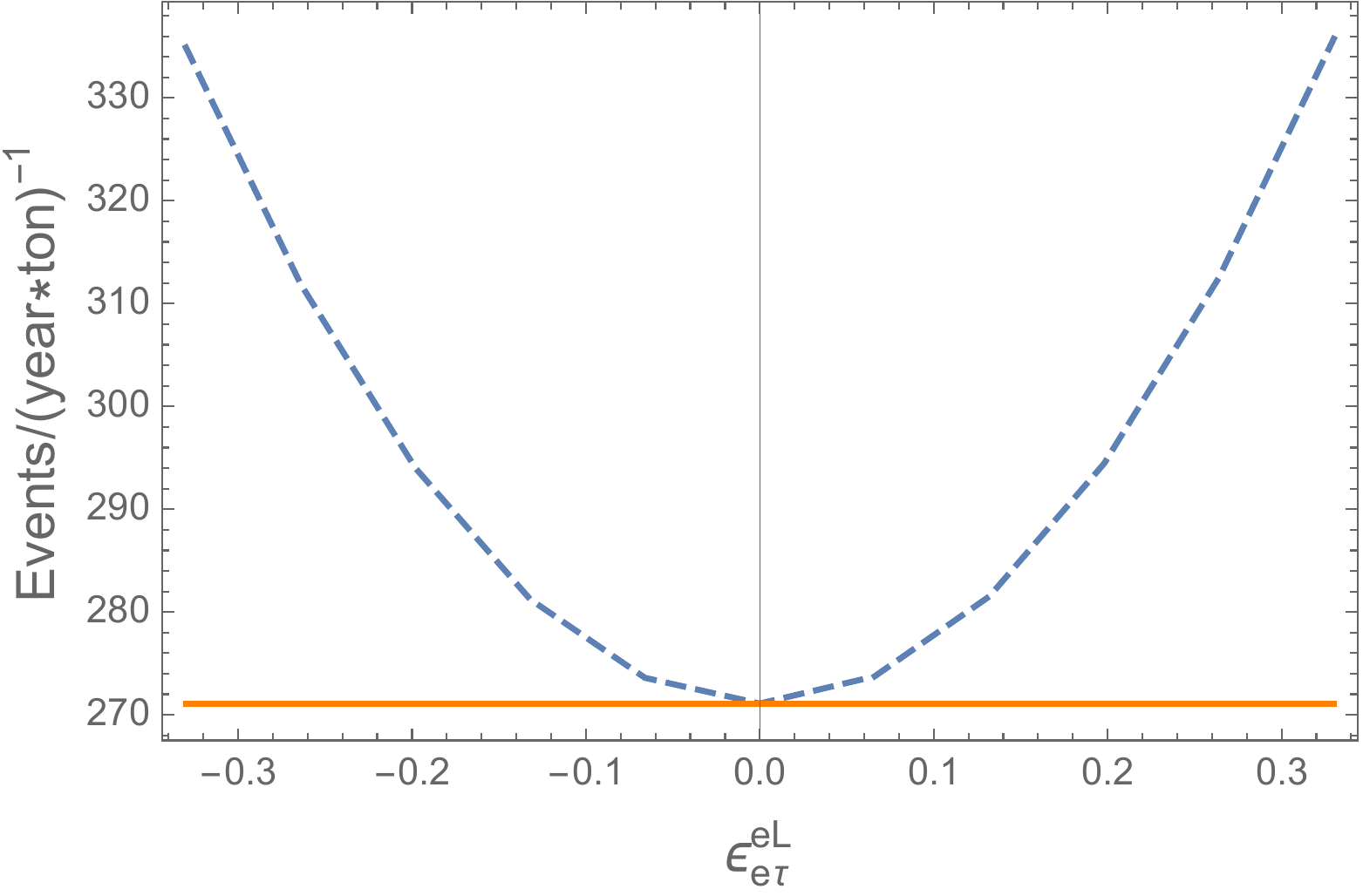} \\
\includegraphics[height=3.5cm]{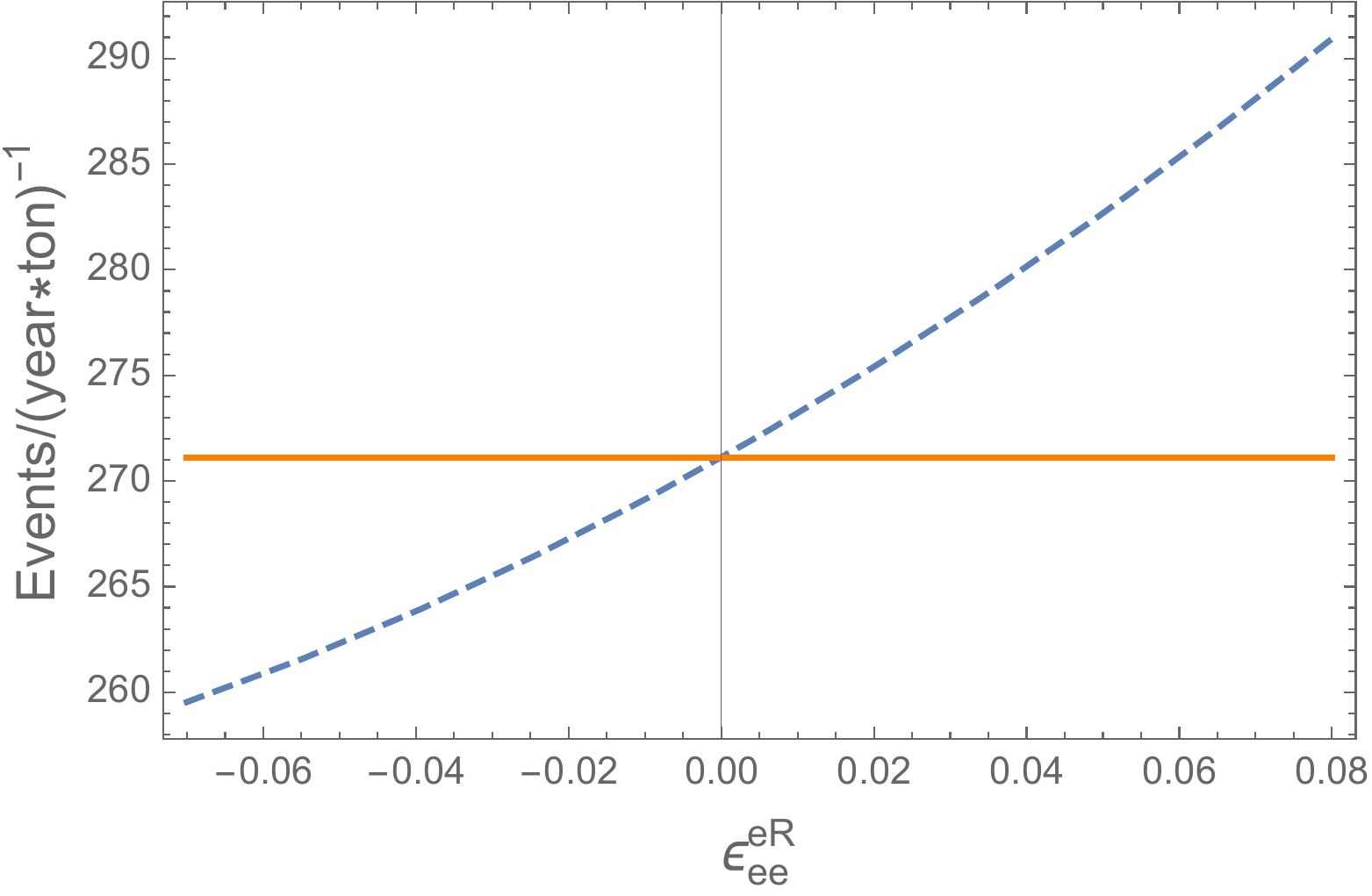} &
\includegraphics[height=3.5cm]{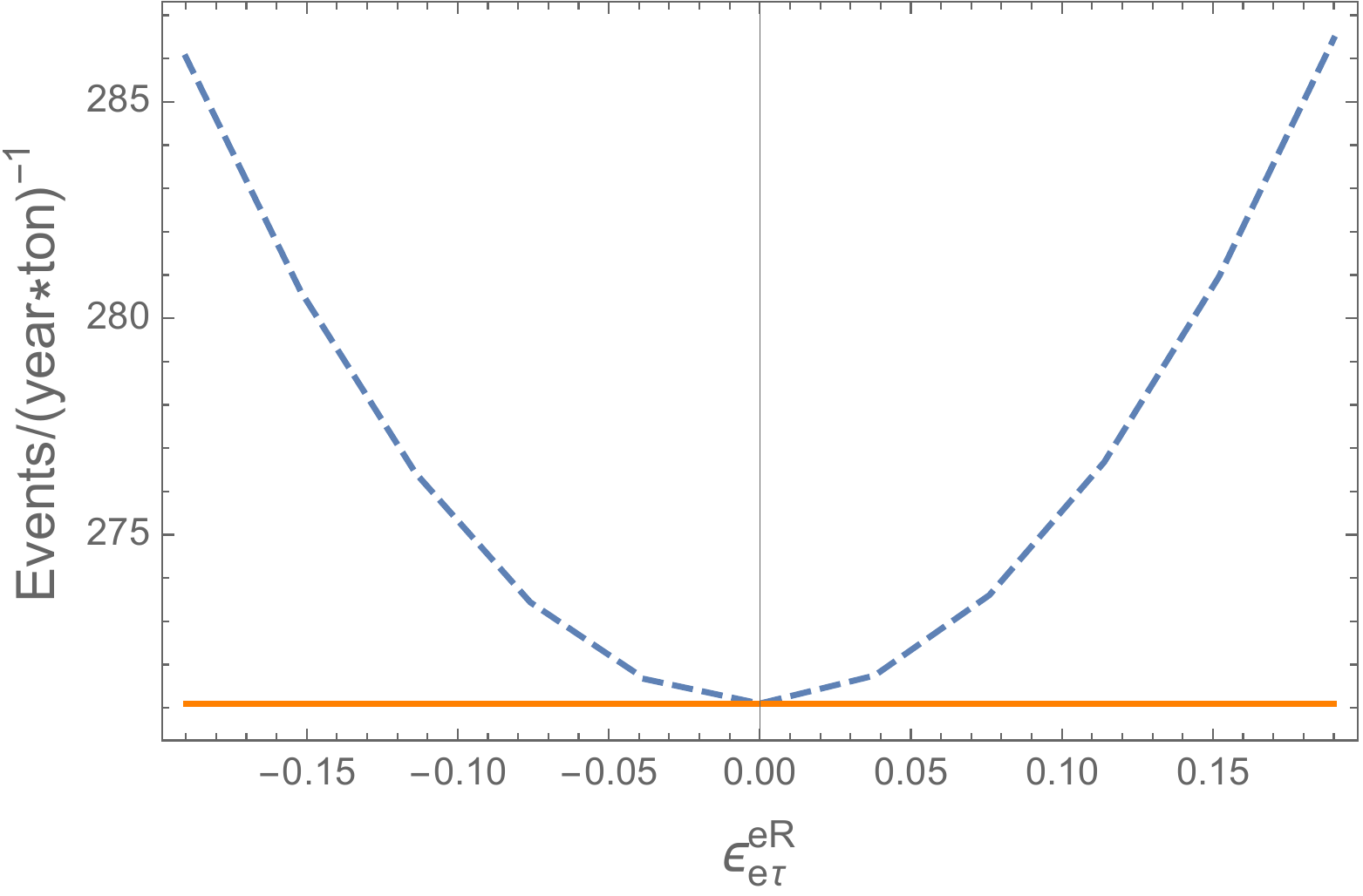} &
\includegraphics[height=3.5cm]{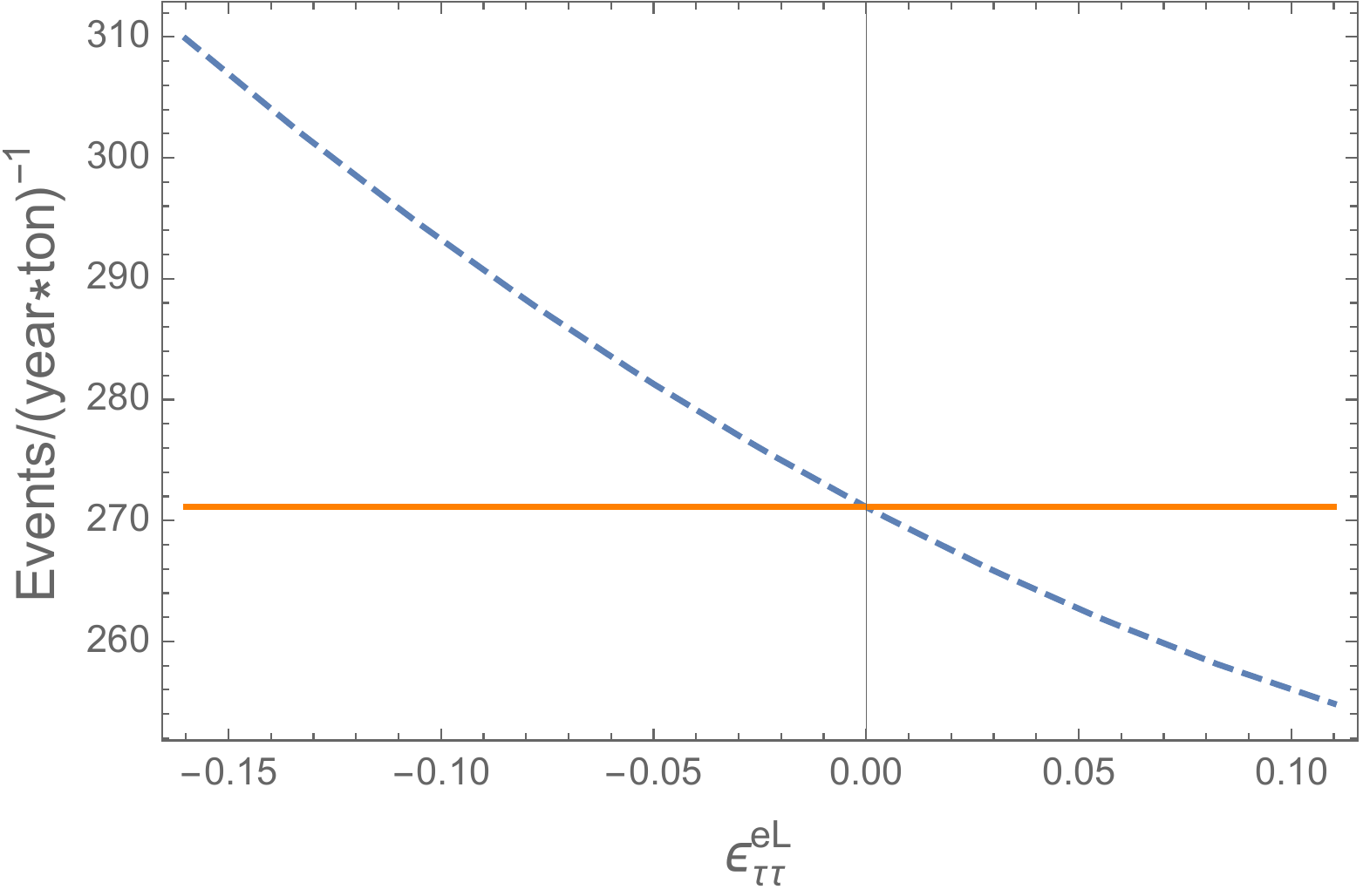} \\
\end{tabular}
\caption{Number of events above an equivalent electron recoil energy threshold of 1 keV as each $\epsilon$ varies over its allowable range (dashed blue curves). The solid orange curve gives the SM contribution.}
\label{fig:epsilons_electron}
\end{figure*}

 Figure~\ref{fig:epsilons_electron} shows how the electron recoil event rate due to elastic scattering, $\nu + e^- \rightarrow \nu + e^-$, is affected as each $\epsilon$ varies over their respective allowed range. The electron recoils are primarily due to low energy $pp$ solar neutrinos, with a $\lesssim$ 10\% contribution from $^7$Be neutrinos. Depending on the value of $\epsilon$, the event rate may either be greater than or less than the corresponding SM event rate. In large regions of parameter space, we find that the existence of NSI parameters can be distinguished from the SM. This can be seen by defining a simple measure of significance as $\chi^{2}=\frac{\left(n_{s}-n_{b}\right)^{2}}{n_{b}}$, where $n_{s}$ is the rate including NSI and $n_b$ is the SM rate. We find that $\epsilon_{e \tau}^{eL}$, $\epsilon_{\tau \tau}^{eL}$ have a significance of $\chi^2 > 4$ for a 1 ton-year exposure.

\par
\begin{figure*}
\begin{tabular}{cccc}
\includegraphics[height=4.2cm]{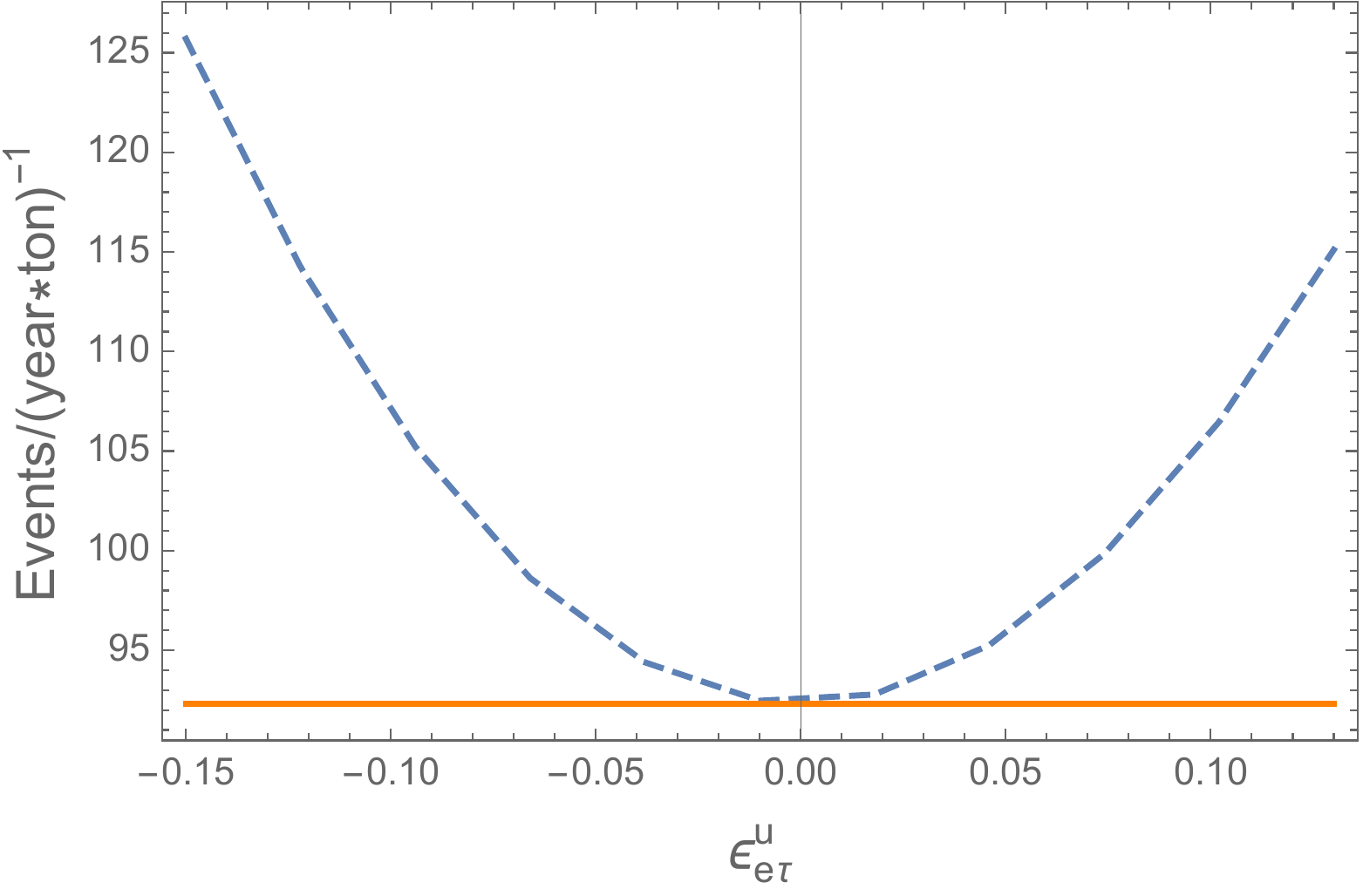} &
\includegraphics[height=4.2cm]{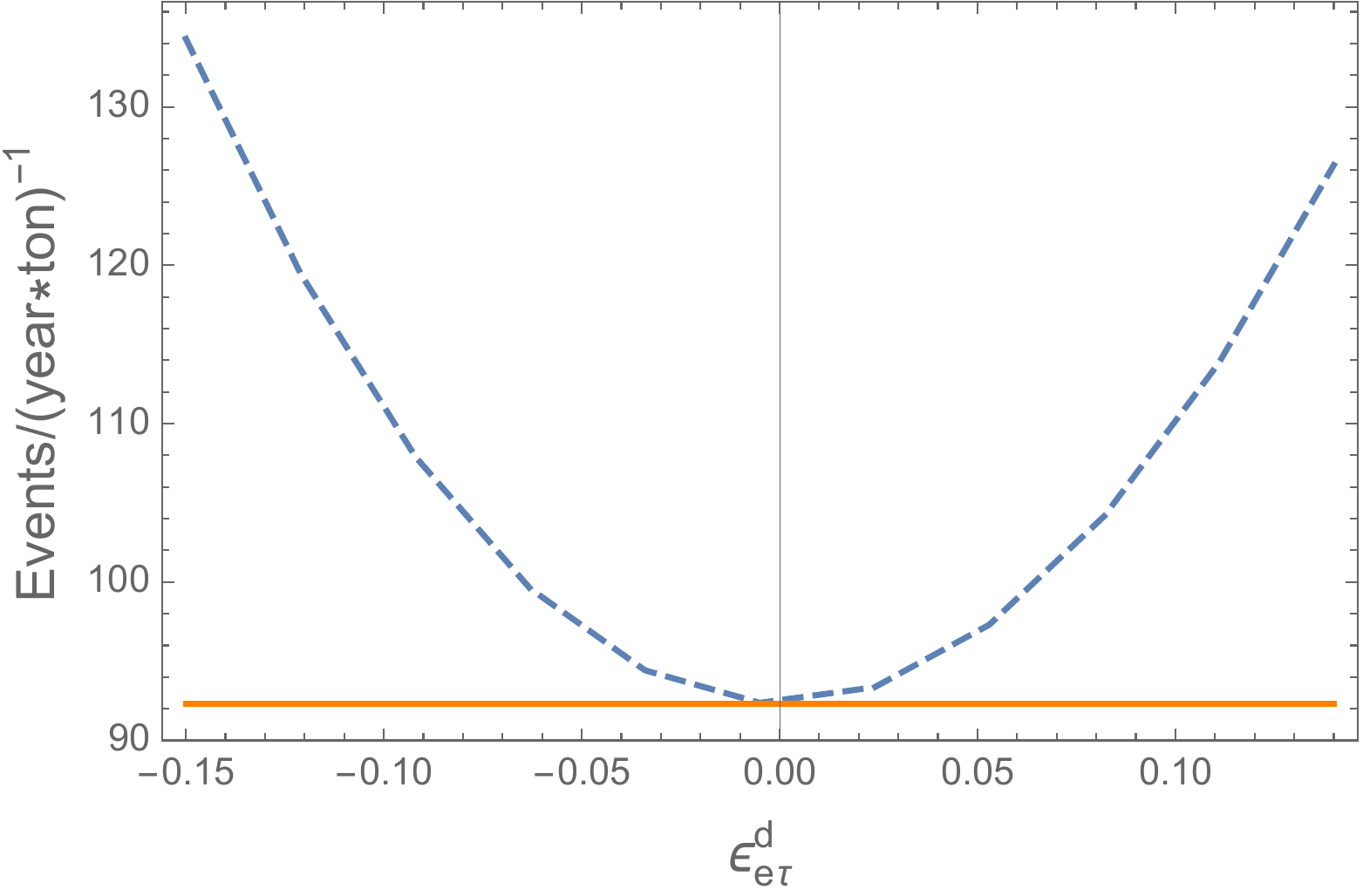} \\
\includegraphics[height=4.2cm]{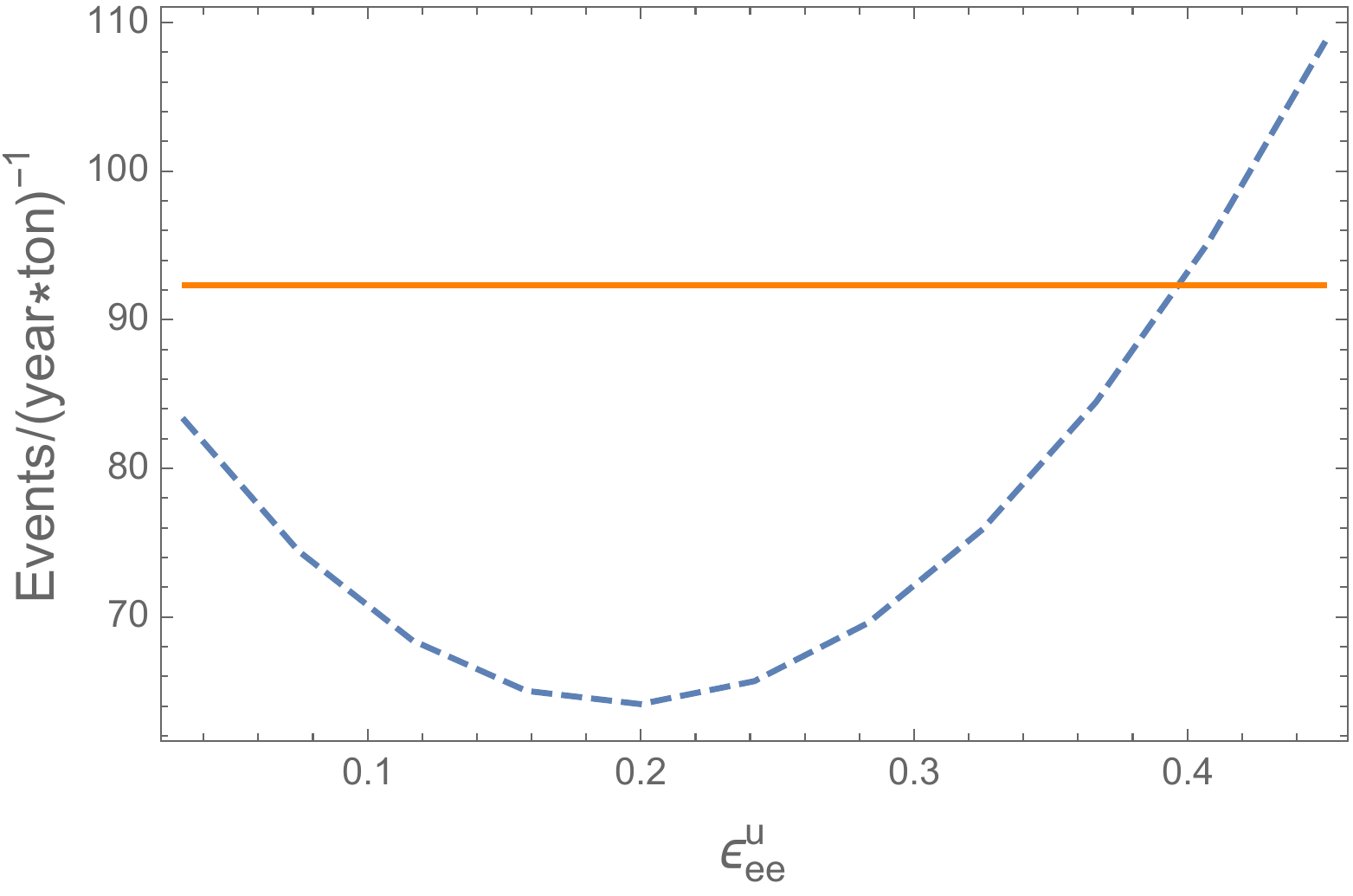} &
\includegraphics[height=4.2cm]{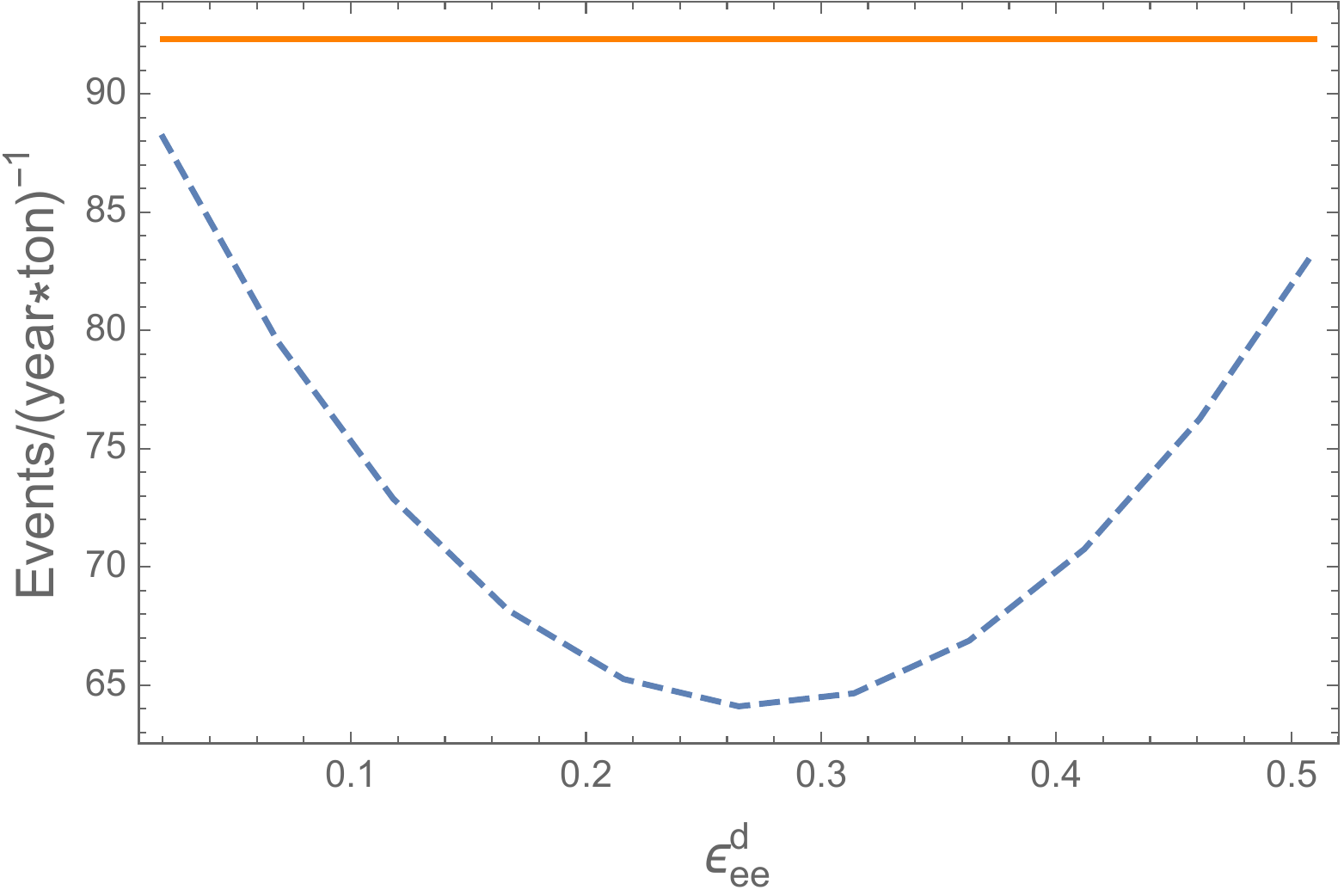} \\
\end{tabular}
\caption{Number of events above a nuclear recoil energy threshold of 1 keV as each $\epsilon$ varies over its allowable range (dashed blue curves). The solid orange curve gives the SM contribution.}
\label{fig:epsilons_nuclear}
\end{figure*}

 Figure~\ref{fig:epsilons_nuclear} shows how the nuclear recoil event rate due to coherent scattering on the xenon nucleus is affected as each $\epsilon$ changes over their respective allowed range. The coherent scattering over the energy regime we consider is due entirely to the high energy, $^8$B solar neutrinos. For nuclear recoils, Figure~\ref{fig:epsilons_nuclear} shows that direct detection experiments are sensitive to $e$ and $\tau$-flavor NSI. As in the case of electron recoils, the event rate may be either enhanced or decreased relative to the SM event rate. As for electron recoils, the existence of NSI parameters can be distinguished from the SM for large regions of parameter space; for example we find that $\epsilon_{ee}^{u}$, $\epsilon_{e \tau}^{u}$, and $\epsilon_{e \tau}^{d}$  have a significance of $\chi^2 > 4$ for a 1 ton-year exposure. In contrast to the case of electron recoils, in order to get an observable effect, the threshold needs to be relatively low, $\sim 1$ keV, as the nuclear form factor suppresses the number of events at high energy drastically. Future xenon experiments are expected to be able to achieve the required threshold energies~\cite{Akerib:2015cja}.

\par The differences between the SM and NSI rates in Figures~\ref{fig:epsilons_electron} and~\ref{fig:epsilons_nuclear} are due to both the effect of NSI in propagation and in detection. In order to disentangle the relative importance of both effects, Table~\ref{tab:msw} shows the event rate compared to the SM rate, with and without the inclusion of NSI term in Hamiltonian of Equation~\ref{eq:hamoltonian}. The first row, $N_{MSW}/N_{SM}$, is the ratio of the respective curves in Figures~\ref{fig:epsilons_electron} and~\ref{fig:epsilons_nuclear} at their $\epsilon$ values of maximal deviation. This is compared to the case in which we do not allow for neutrino flavor transformations from their production in the sun to their detection; the event rate in this case is defined as $N_{noMSW}$. Instances in which the ratios $N_{noMSW}/N_{SM}$ and $N_{MSW}/N_{SM}$ are close together implies that the largest impact of NSI comes from the detection cross section in Equation~\ref{eq:crossSect}. Where these ratios are very different, such as with $\epsilon_{ee}^u$ and $\epsilon_{ee}^d$, the dominant effect of NSI comes from matter-induced transformations.

\par
\begin{figure}
\begin{tabular}{cc}
\includegraphics[height=4.2cm]{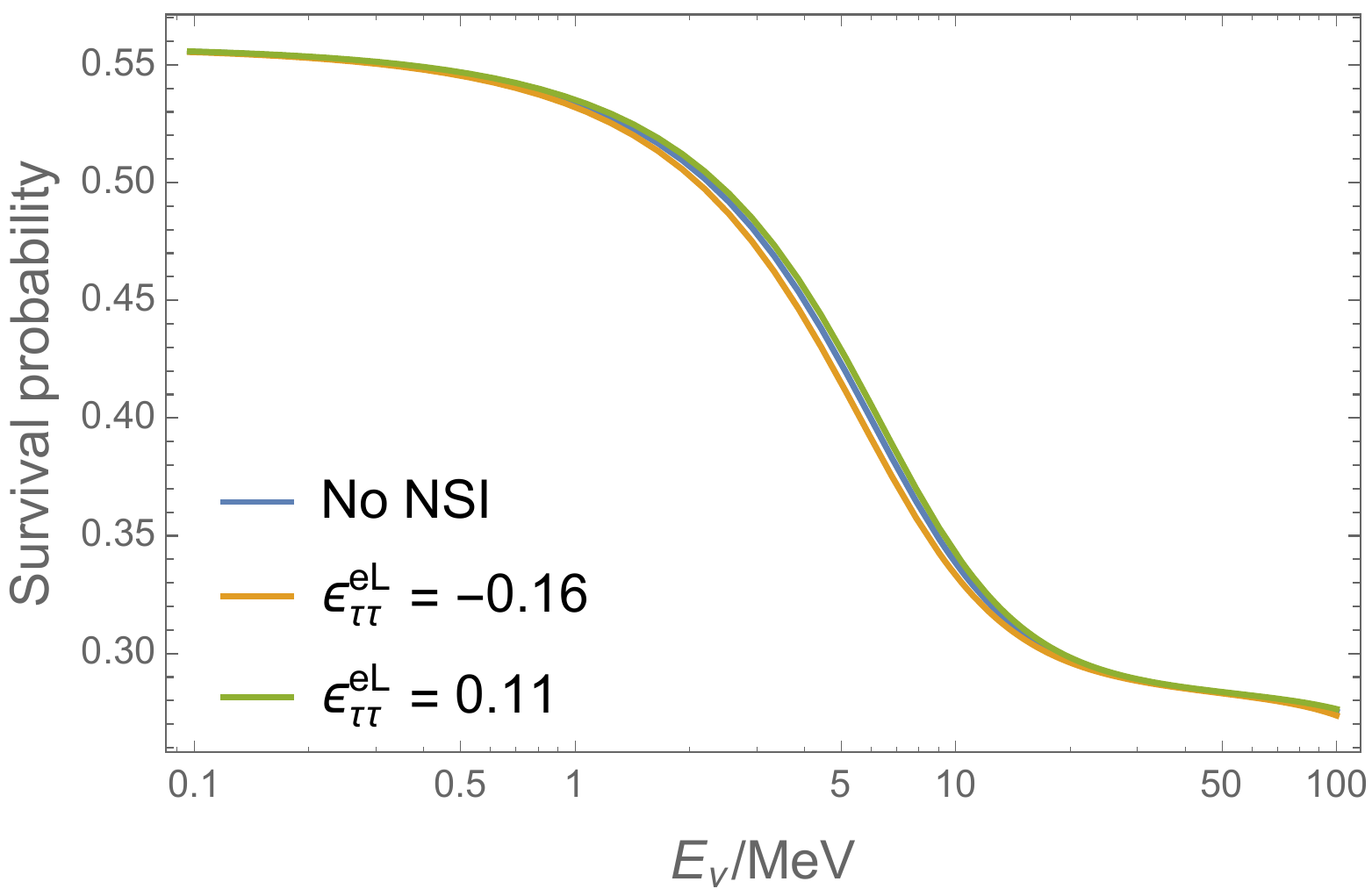} &
\includegraphics[height=4.2cm]{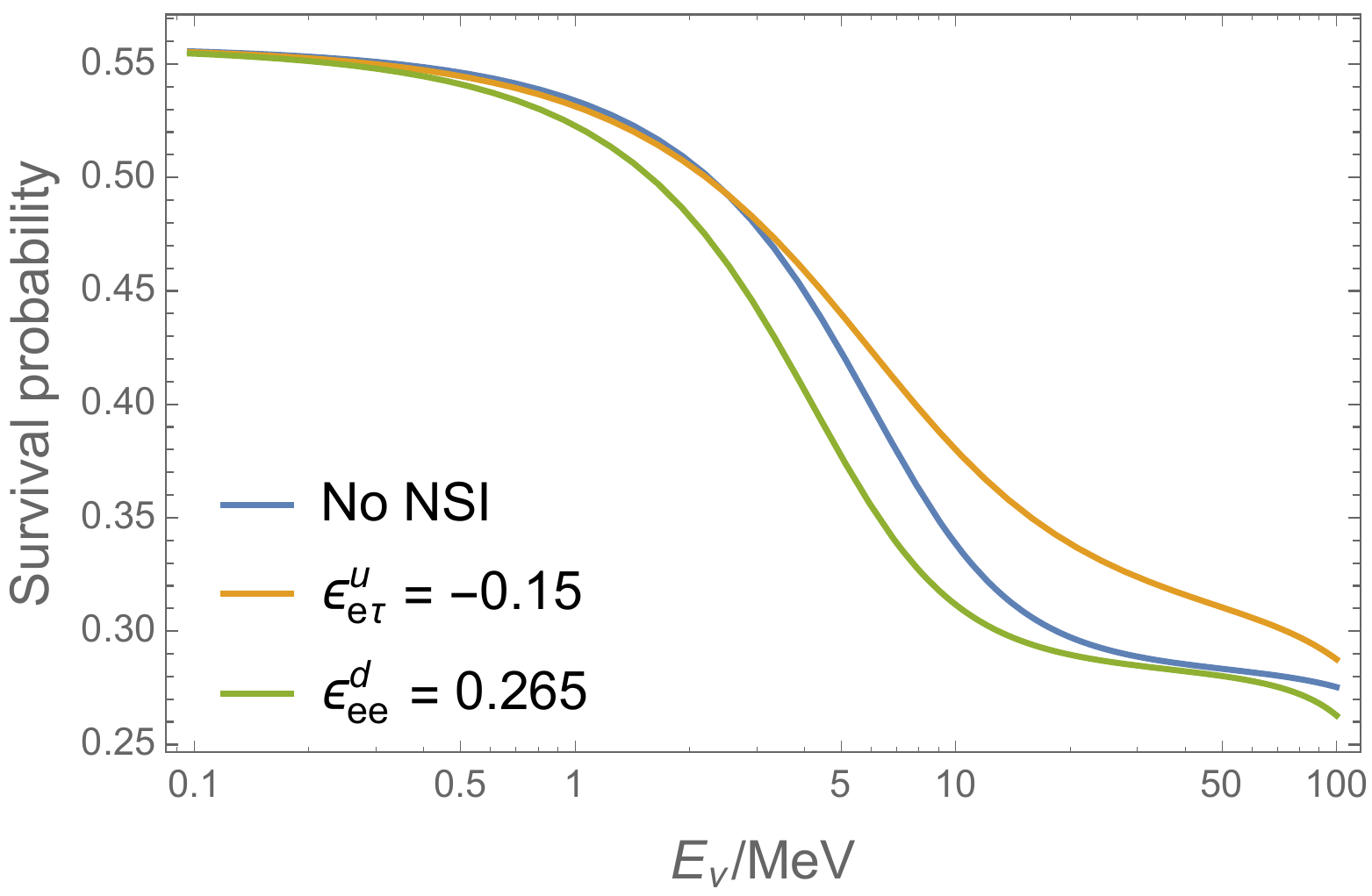} \\
\end{tabular}
\caption{Electron neutrino survival probability for the SM (blue) compared to cases in which the NSI give significant deviations from the SM. The left panel shows $\epsilon$'s which give deviation from the SM for electron recoils, and the right panel shows $\epsilon$'s which give deviation from the SM for nuclear recoils.}
\label{fig:survival}
\end{figure}

 To further illustrate this point for a couple of examples, we can examine the electron neutrino survival probability with NSI included. Figure~\ref{fig:survival} shows the survival probability for several $\epsilon$'s which show significant deviation from the SM, for both electron and nuclear recoils. In both cases, maximal deviations between the NSI and SM curves arise in the transition to the matter-dominated regime and in the matter dominated region, while the vacuum-dominated regime at low energy is unaffected.

\begin{table}
\scalebox{0.82}{
\begin{tabular}{|c|c|c|c|c|c|c|c|c|c|c|}
\hline
 & $\epsilon_{ee}^{eL}=0.052$ & $\epsilon_{e\mu}^{eL}=0.13$ & $\epsilon_{e\tau}^{eL}=0.33$ & $\epsilon_{ee}^{eR}=0.08$ & $\epsilon_{e\tau}^{eR}=0.19$ & $\epsilon_{\tau\tau}^{eL}=-0.16$ & $\epsilon_{ee}^{u}=0.2$ & $\epsilon_{e\tau}^{u}=-0.15$ & $\epsilon_{e\tau}^{d}=-0.15$ & $\epsilon_{ee}^{d}=0.26$\tabularnewline
\hline
\hline
$\frac{N}{N_{SM}}$ & $1.10$ & $1.03$ & 1.23 & $1.07$ & $1.05$ & $1.14$ & $0.69$ & $1.36$ & $1.45$ & $0.69$\tabularnewline
\hline
$\frac{N}{N_{SM}}$ & $1.12$ & $1.03$ & $1.31$ & $1.01$ & $1.07$ & $1.08$ & $0.67$ & $1.34$ & $1.43$ & $0.75$\tabularnewline
\hline
$\frac{N}{N_{SM}}$ & $1.13$ & $1.05$ & $1.32$ & $1.02$ & $1.08$ & $1.09$ & $0.44$ & $1.43$ & $1.55$ & $0.56$\tabularnewline
\hline
$\frac{N}{N_{SM}}$ & $1.69$ & $1.52$ & $1.79$ & $1.51$ & $1.55$ & $1.47$ & $4\times10^{-5}$ & $1.56$ & $19.07$ & $5\times10^{-3}$\tabularnewline
\hline
\end{tabular}
}
\caption{Each row gives the ratio of the number of events for a given $\epsilon$, with and without the inclusion of neutrino transformations in Equation~\ref{eq:hamoltonian}. $N_{SM}$ is the number of events in the SM, without NSI, $N$ is the number of events predicted by substracting NSI-MSW, MSW and entire oscillation from Equation~\ref{eq:hamoltonian} correspondly for each row.  For each column, the $\epsilon$ is chosen to maximize the difference from SM.}
\label{tab:msw}
\end{table}

\par
\begin{figure}
\begin{tabular}{c}
\includegraphics[height=5.0cm]{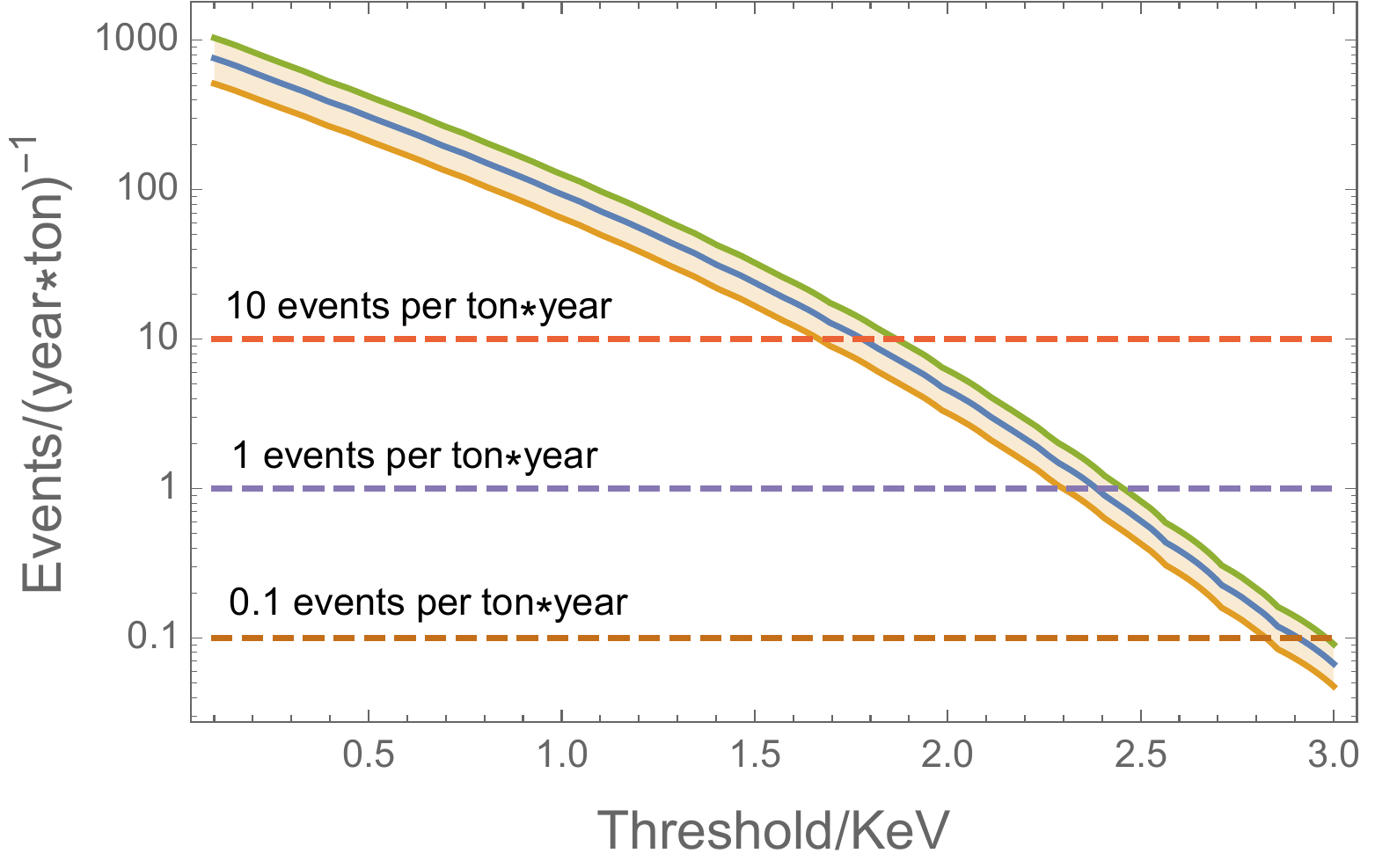} \\
\end{tabular}
\caption{Event rate as a function of energy threshold for different NSI models. Horizontal lines indicate where the event rate falls below one per year for, from top to bottom, 0.1, 1 and 10 ton detectors.}
\label{fig:threshold}
\end{figure}

 The aforementioned results clearly indicate that NSI will affect future low-mass dark matter searches. Previous studies have used a specific statistical criteria, i.e. a discovery limit~\cite{Billard:2013qya,Dent:2016iht}, to quantify how the dark matter sensitivity scales as a function of detector exposure. For simplicity, here we just consider dark matter searches to be significantly impacted when the number of neutrino events above a given nuclear energy threshold exceeds one, for a given detector exposure. Figure~\ref{fig:threshold} shows how this event rate depends on energy threshold, for NSI parameters which give a maximal deviation from the SM. This clearly indicates how the neutrino floor may ultimately be either raised or lowered if NSI are allowed.

\par
\begin{figure}
\begin{tabular}{c}
\includegraphics[height=5.0cm]{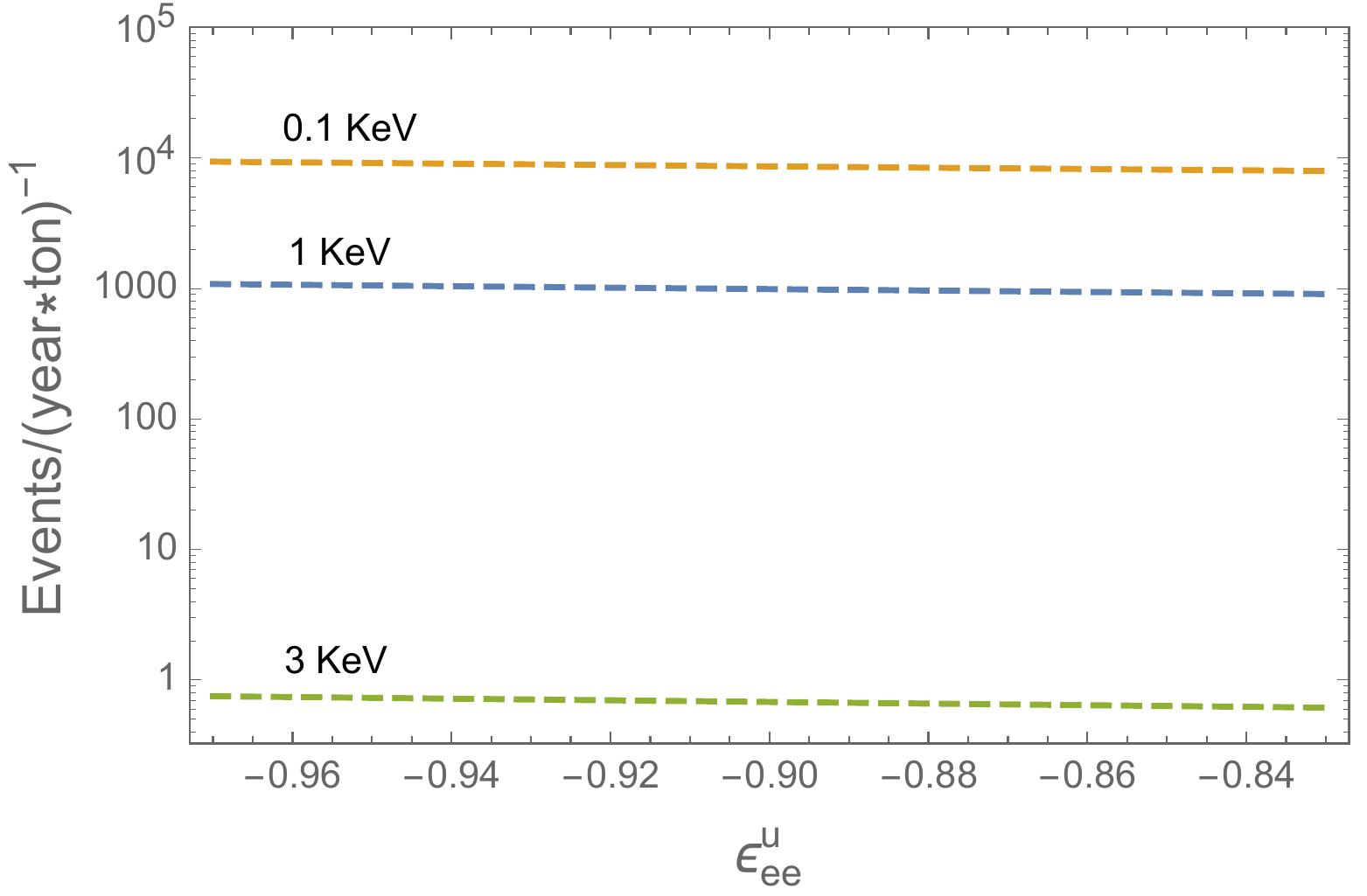} \\
\end{tabular}
\caption{Total number of events above an indicated nuclear threshold energy for the LMA-d solution with $\sin^2 \theta_{12} = 0.7$ \cite{Miranda:2004nb}, for three different threshold energies.}
\label{fig:lmad}
\end{figure}

 Finally, we note that when NSI are allowed, a ``dark side" solution for the LMA appears, characterized by $\theta_{12} > 45^\circ$ (LMA-d)~\cite{Gonzalez-Garcia:2013usa}. In Figure~\ref{fig:lmad}, we show that this solution can be discovered in direct detection experiments for threshold energies of 1 keV. Thus forthcoming direct detection experiments have a novel and unique discovery sensitivity to the entire region of the LMA-d solution.

\section{Discussion and Conclusion}
\par In this paper we have explored the impact of non-standard neutrino interactions on the neutrino background in dark matter detectors, focusing specifically on solar neutrinos. We have shown that due to both matter-induced transformations in propagation and changes in the cross section, the event rate can be significantly enhanced or decreased relative to the SM prediction. This change may be observable in forthcoming detectors.

\par Two important and general conclusions follow from our results. First, direct dark matter searches are able to probe NSI parameter space that cannot be probed by current neutrino experiments, implying that direct dark matter searches have an important role in searches for beyond the Standard Model physics through the neutrino sector. The elastic scattering and coherent scattering channels are complementary ~\cite{Coloma:2017egw}, probing NSI in different regimes of parameter space. Second, the uncertainty in NSI parameters constitute an additional uncertainty to the neutrino background in future searches for low mass dark matter.

\par While the focus of this paper has been on solar neutrinos, it will be straightforward to extend this analysis to both atmospheric and diffuse supernova neutrinos, which will constitute a background for dark matter experiments at higher dark matter mass, $\gtrsim 10$ GeV. However, it is likely that the atmospheric background and NSI contributions to the background cannot be reached soon. The implications for dark matter searches over a wide range of mass scales will be further explored in forthcoming analyses.

\bigskip

{\bf Acknowledgements}:
BD acknowledges support from DOE Grant DE-FG02-13ER42020. LES acknowledges support from NSF grant PHY-1522717. JWW acknowledges support from NSF grant PHY-1521105. We thank Matthew Breeding and Derek Johnson for contributions to code for characterization of the Solar neutrino flux.  We thank James Dent, Jayden Newstead, Ian Shoemaker, Osamu Yasuda for useful discussions throughout the course of this work.


\begin{thebibliography}{99}

\bibitem{Akerib:2015rjg}
  D.~S.~Akerib {\it et al.} [LUX Collaboration],
  Phys.\ Rev.\ Lett.\  {\bf 116}, no. 16, 161301 (2016)
  [arXiv:1512.03506 [astro-ph.CO]].

\bibitem{Aprile:2012nq}
  E.~Aprile {\it et al.} [XENON100 Collaboration],
  Phys.\ Rev.\ Lett.\  {\bf 109}, 181301 (2012)
  [arXiv:1207.5988 [astro-ph.CO]].

\bibitem{Agnese:2015nto}
  R.~Agnese {\it et al.} [SuperCDMS Collaboration],
  Phys.\ Rev.\ Lett.\  {\bf 116}, no. 7, 071301 (2016)
  [arXiv:1509.02448 [astro-ph.CO]].

\bibitem{Angloher:2015ewa}
  G.~Angloher {\it et al.} [CRESST Collaboration],
  Eur.\ Phys.\ J.\ C {\bf 76}, no. 1, 25 (2016)
  [arXiv:1509.01515 [astro-ph.CO]].

\bibitem{Armengaud:2016cvl}
  E.~Armengaud {\it et al.} [EDELWEISS Collaboration],
  arXiv:1603.05120 [astro-ph.CO].

\bibitem{Monroe:2007xp}
  J.~Monroe and P.~Fisher,
  Phys.\ Rev.\ D {\bf 76}, 033007 (2007)
  [arXiv:0706.3019 [astro-ph]].

\bibitem{Strigari:2009bq}
  L.~E.~Strigari,
  New J.\ Phys.\  {\bf 11}, 105011 (2009)
  [arXiv:0903.3630 [astro-ph.CO]].

\bibitem{Billard:2013qya}
  J.~Billard, L.~Strigari and E.~Figueroa-Feliciano,
  Phys.\ Rev.\ D {\bf 89}, no. 2, 023524 (2014)
  [arXiv:1307.5458 [hep-ph]].

\bibitem{Ruppin:2014bra}
  F.~Ruppin, J.~Billard, E.~Figueroa-Feliciano and L.~Strigari,
  Phys.\ Rev.\ D {\bf 90}, no. 8, 083510 (2014)
  [arXiv:1408.3581 [hep-ph]].

\bibitem{Pospelov:2011ha}
  M.~Pospelov,
  Phys.\ Rev.\ D {\bf 84}, 085008 (2011)
  [arXiv:1103.3261 [hep-ph]].

\bibitem{Harnik:2012ni}
  R.~Harnik, J.~Kopp and P.~A.~N.~Machado,
  JCAP {\bf 1207}, 026 (2012)
  [arXiv:1202.6073 [hep-ph]].

\bibitem{Billard:2014yka}
  J.~Billard, L.~E.~Strigari and E.~Figueroa-Feliciano,
  Phys.\ Rev.\ D {\bf 91}, no. 9, 095023 (2015)
  [arXiv:1409.0050 [astro-ph.CO]].

\bibitem{Cerdeno:2016sfi}
  D.~G.~CerdeÃ±o, M.~Fairbairn, T.~Jubb, P.~A.~N.~Machado, A.~C.~Vincent and C.~BÅhm,
  JHEP {\bf 1605}, 118 (2016)
  Erratum: [JHEP {\bf 1609}, 048 (2016)]
  [arXiv:1604.01025 [hep-ph]].

\bibitem{Dent:2016wcr}
  J.~B.~Dent, B.~Dutta, S.~Liao, J.~L.~Newstead, L.~E.~Strigari and J.~W.~Walker,
  arXiv:1612.06350 [hep-ph].

\bibitem{Kosmas:2017zbh}
  T.~S.~Kosmas, D.~K.~Papoulias, M.~Tortola and J.~W.~F.~Valle,
  arXiv:1703.00054 [hep-ph].

\bibitem{Lindner:2016wff}
  M.~Lindner, W.~Rodejohann and X.~J.~Xu,
  JHEP {\bf 1703}, 097 (2017)
  [arXiv:1612.04150 [hep-ph]].

\bibitem{Shoemaker:2017lzs}
  I.~M.~Shoemaker,
  arXiv:1703.05774 [hep-ph].

\bibitem{Ohlsson:2012kf}
  T.~Ohlsson,
  Rept.\ Prog.\ Phys.\  {\bf 76}, 044201 (2013)
  [arXiv:1209.2710 [hep-ph]].

\bibitem{Miranda:2015dra}
  O.~G.~Miranda and H.~Nunokawa,
  New J.\ Phys.\  {\bf 17}, no. 9, 095002 (2015)
  [arXiv:1505.06254 [hep-ph]].

\bibitem{Maltoni:2015kca}
  M.~Maltoni and A.~Y.~Smirnov,
  Eur.\ Phys.\ J.\ A {\bf 52}, no. 4, 87 (2016)
  [arXiv:1507.05287 [hep-ph]].

\bibitem{Fukasawa:2016nwn}
  S.~Fukasawa and O.~Yasuda,
  Nucl.\ Phys.\ B {\bf 914}, 99 (2017)
  [arXiv:1608.05897 [hep-ph]].

\bibitem{deGouvea:2015ndi}
  A.~de Gouvêa and K.~J.~Kelly,
  Nucl.\ Phys.\ B {\bf 908}, 318 (2016)
  [arXiv:1511.05562 [hep-ph]].

\bibitem{Friedland:2004ah}
  A.~Friedland, C.~Lunardini and M.~Maltoni,
  Phys.\ Rev.\ D {\bf 70}, 111301 (2004)
  [hep-ph/0408264].

\bibitem{Bolanos:2008km}
  A.~Bolanos, O.~G.~Miranda, A.~Palazzo, M.~A.~Tortola and J.~W.~F.~Valle,
  Phys.\ Rev.\ D {\bf 79}, 113012 (2009)
  [arXiv:0812.4417 [hep-ph]].

\bibitem{Agarwalla:2012wf}
  S.~K.~Agarwalla, F.~Lombardi and T.~Takeuchi,
  JHEP {\bf 1212}, 079 (2012)
  [arXiv:1207.3492 [hep-ph]].

\bibitem{Stapleford:2016jgz}
  C.~J.~Stapleford, D.~J.~Vnnen, J.~P.~Kneller, G.~C.~McLaughlin and B.~T.~Shapiro,
  Phys.\ Rev.\ D {\bf 94}, no. 9, 093007 (2016)
  [arXiv:1605.04903 [hep-ph]].

\bibitem{Gonzalez-Garcia:2016gpq}
  M.~C.~Gonzalez-Garcia, M.~Maltoni, I.~Martinez-Soler and N.~Song,
  Astropart.\ Phys.\  {\bf 84}, 15 (2016)
  [arXiv:1605.08055 [hep-ph]].

\bibitem{Aprile:2015uzo}
  E.~Aprile {\it et al.} [XENON Collaboration],
  JCAP {\bf 1604}, no. 04, 027 (2016)
  [arXiv:1512.07501 [physics.ins-det]].

\bibitem{Mount:2017qzi}
  B.~J.~Mount {\it et al.},
  arXiv:1703.09144 [physics.ins-det].

\bibitem{Barranco:2005yy}
  J.~Barranco, O.~G.~Miranda and T.~I.~Rashba,
  JHEP {\bf 0512}, 021 (2005)
  [hep-ph/0508299].

\bibitem{Lewin:1995rx}
  J.~D.~Lewin and P.~F.~Smith,
  Astropart.\ Phys.\  {\bf 6}, 87 (1996).

\bibitem{Bahcall:2004pz}
  J.~N.~Bahcall, A.~M.~Serenelli and S.~Basu,
  Astrophys.\ J.\  {\bf 621}, L85 (2005)
  doi:10.1086/428929
  [astro-ph/0412440].

\bibitem{Kuo:1989qe}
  T.~K.~Kuo and J.~T.~Pantaleone,
  Rev.\ Mod.\ Phys.\  {\bf 61}, 937 (1989).

\bibitem{Friedland:2004pp}
  A.~Friedland, C.~Lunardini and C.~Pena-Garay,
  Phys.\ Lett.\ B {\bf 594}, 347 (2004)
  [hep-ph/0402266].

\bibitem{Gonzalez-Garcia:2013usa}
  M.~C.~Gonzalez-Garcia and M.~Maltoni,
  JHEP {\bf 1309}, 152 (2013)
  [arXiv:1307.3092 [hep-ph]].

\bibitem{Coloma:2017egw}
  P.~Coloma, P.~B.~Denton, M.~C.~Gonzalez-Garcia, M.~Maltoni and T.~Schwetz,
  arXiv:1701.04828 [hep-ph].

\bibitem{Olive:2016xmw}
  C.~Patrignani {\it et al.} [Particle Data Group],
  Chin.\ Phys.\ C {\bf 40}, no. 10, 100001 (2016).

\bibitem{Antonelli:2012qu}
  V.~Antonelli, L.~Miramonti, C.~Pena Garay and A.~Serenelli,
  Adv.\ High Energy Phys.\  {\bf 2013}, 351926 (2013)
  [arXiv:1208.1356 [hep-ex]].

\bibitem{Robertson:2012ib}
  W.~C.~Haxton, R.~G.~Hamish Robertson and A.~M.~Serenelli,
  Ann.\ Rev.\ Astron.\ Astrophys.\  {\bf 51}, 21 (2013)
  [arXiv:1208.5723 [astro-ph.SR]].

\bibitem{Akerib:2015cja}
  D.~S.~Akerib {\it et al.} [LZ Collaboration],
  arXiv:1509.02910 [physics.ins-det].

\bibitem{Dent:2016iht}
  J.~B.~Dent, B.~Dutta, J.~L.~Newstead and L.~E.~Strigari,
  Phys.\ Rev.\ D {\bf 93}, no. 7, 075018 (2016)
  [arXiv:1602.05300 [hep-ph]].

\bibitem{Miranda:2004nb}
    O.~G.~Miranda, M.~A.~Tortola and J.~W.~F.~Valle,
    JHEP {\bf 0610}, 008 (2006)
    doi:10.1088/1126-6708/2006/10/008
    [hep-ph/0406280].

\end{thebibliography}
\end{document}